\documentclass[journal,onecolumn]{IEEEtran}
\usepackage{epsfig}
\usepackage{graphicx}
\usepackage{graphics}
\usepackage{epstopdf} 
\usepackage{tikz}
\usetikzlibrary{calc,shapes.geometric}
\usetikzlibrary{arrows,snakes,backgrounds}

\ifCLASSINFOpdf
\else
\fi

\usepackage[cmex10]{amsmath}

\def\clock{\n=\time \divide\n 60
  \m=-\n \multiply\m 60 \advance\m \time
  \ifnum \n>12 \advance\n -12 \fi
   \number\n.\twodigits\m~\ampm\time}
\def\ampm#1{\ifnum #1< 720 am\else pm\fi}
\def\twodigits#1{\ifnum #1<10 0\fi \number#1}

\usepackage{epsfig}

\def\nexto{\kern -0.54em}

\def\prob{{\rm {I\ \nexto P}}}

\def\G{{\cal G}}
\def\S{{\cal S}}
\def\C{{\cal C}}
\def\s{{\cal K}}
\def\I{{\cal I}}
\def\Z{{\cal Z}}
\def\T{{\cal T}}

\newcount\m \newcount\n
\def\clock{\n=\time \divide\n 60
  \m=-\n \multiply\m 60 \advance\m \time
  \ifnum \n>12 \advance\n -12 \fi
   \number\n.\twodigits\m~\ampm\time}
\def\ampm#1{\ifnum #1< 720 am\else pm\fi}
\def\twodigits#1{\ifnum #1<10 0\fi \number#1}

\begin{document}

\title{Approximation of the Cell Under Test in Sliding Window Detection Processes}
\author{Graham  V. Weinberg  \\ (Draft created at \clock)\\
 }
\maketitle

\markboth{Approximation of the Cell Under Test \today}%
{}

\begin{abstract}
Analysis of sliding window detection detection processes requires careful consideration of the cell under test, which is an amplitude squared measurement of the signal plus clutter in the complex domain. Some authors have suggested that there is sufficient merit in the approximation of the cell under test by a distributional model similar to that assumed for the clutter distribution. Under such an assumption, the development of Neyman-Pearson detectors is facilitated.
The purpose of the current paper is to demonstrate, in a modern maritime surveillance radar context, that such an approximation is invalid. 
\end{abstract}


\section{Introduction}
Sliding window detectors assume the existence of a series of clutter measurements, from which a single measurement of the clutter level is taken. This is then compared with a cell under test (CUT), which is being tested for the presence of a target in clutter. The measurement of clutter level is then normalised in such a way that the probability of false alarm (Pfa) can be stabilised adaptively. In a maritime surveillance radar context, one could be examining returns from the sea surface. Clutter measurements are returns from the sea surface in the absence of a target. The cell under test, in the presence of an artefact of interest,  is a return which is taken independently and could consist of the signature of a boat sitting on the sea surface, for example.
The analysis of these detection processes began with the pioneering work in \cite{finn}; a modern account can be found in \cite{weinbergbook}.

The problem of interest, in the current paper, is analysis of the CUT in the presence of a target. Some recent studies have suggested that the CUT could be approximated by the same distribution used for the clutter, but with different distributional parameters. This is certainly the case in the context of \cite{finn}. Such an approach, in more modern situations, is suggested in \cite{siddiq}, who apply an approximation from \cite{dongy} for the sum of a series of Weibull distributed variables, to facilitate the derivation of the single measurement of clutter. In addition to this, an approximation is applied for the CUT, relative to the additivity associated with Rayleigh variates. A second, and more recent treatment, can be found in \cite{persson}, who suggests in the case of more modern X-band maritime surveillance radar, the return in the presence of a target, can be modelled by a Pareto distribution, which is also used for the clutter measurements. The difference is that the Pareto parameters are modified to account for the presence of a target.

This paper will show that this is a somewhat invalid assumption in the context of modern maritime surveillance radar.

\section{Problem Specification}
The CUT consists of a single complex clutter measurement added to a complex target model. For the puposes of simplicity, it is assumed that the target model is bivariate Gaussian. In particular, it is assumed that its in-phase and quadrature components are independent, with zero mean and variance the reciprocal of $2\lambda$, for some $\lambda > 0$. This means that the signal power is of the order the reciprocal of $\lambda$. This signal is denoted $\S$ througout.
The complex clutter return is assumed to result from a compound Gaussian process with inverse gamma texture, which yields the Pareto Type II clutter model in the intensity realm \cite{weinbergbook}. Hence the clutter takes the form $\C = \s \G$, where $\G$ is also a bivariate Gaussian process, with independent in-phase and quadrature components, with zero mean and variance the reciprocal of $2\mu$. The univariate process $\s$, called the speckle, is assumed to be an independent inverse gamma distributed random variable, with density
\begin{equation}
f_{\s}(t) = \frac{\beta^\alpha}{\Gamma(\alpha)} t^{-2\alpha-1} e^{-\beta t^{-2}}, \label{texture}
\end{equation}
for $t > 0$, where $\Gamma$ is the gamma function. It can be shown that the intensity random variable $\I = |\C|^2$ has the Pareto Type II density
\begin{equation}
f_{\I}(t) = \frac{\alpha \beta^\alpha}{( t + \beta)^{\alpha+1}}, \label{pardens}
\end{equation}
and corresponding distribution function
\begin{equation}
F_{\I}(t) = \prob(\I \leq t) = 1 - \left( \frac{\beta}{t + \beta}\right)^\alpha, \label{parcdf}
\end{equation}
with $t \geq 0$. In the above, parameter $\alpha > 0$ is the shape, while $\beta > 0$ is the scale.

The CUT, in the presence of a target, is simply $\Z = \left|\S +\C\right|^2 $, and the problem of interest is to determine its distribution. 
The approach to be examined is the assumption that there are cases where the distribution of $\Z$ is also Pareto distributed, but with modified shape and scale parameters.

\section{An Approximation for the CUT}
For the scenario specified in the previous section, it is shown in \cite{weinberg13} that the distribution function\footnote{This corrects a typesetting error in \cite{weinberg13}, where a redundant 2 appears in the expression.} of $\Z$ is given by
\begin{equation}
F_{\Z(t)} = 1 - \frac{\beta^\alpha}{\Gamma(\alpha)} \int_0^\infty u^{\alpha-1} e^{-\beta u} e^{-\frac{t}{ \frac{1}{\lambda} + \frac{1}{\mu u}}} du,
\label{cdf1}
\end{equation}
where $t \geq 0$. It is of interest to determine whether there are any conditions on the parameters $\alpha, \beta, \lambda$ and $\mu$ that result in \eqref{cdf1} being approximately Pareto distributed, with a distribution function of the form \eqref{parcdf}.

Examination of the integrand in \eqref{cdf1} suggests it is necessary to examine approximations for its third term. Towards this objective, define a function
\begin{equation}
g(u) = \left[ \frac{1}{\lambda} + \frac{1}{\mu}u^{-1}\right]^{-1} = \frac{\lambda \mu u}{\mu u + \lambda}. \label{gfun1}
\end{equation}
Then the term under consideration can be written 
\begin{equation}
\T = e ^{-g(u) t}. \label{gfun2}
\end{equation}
One requires a reasonable approximation for \eqref{gfun1}, so that when applied to \eqref{gfun2}, and subsequently to \eqref{cdf1}, it becomes possible to evaluate this integral.

Observe that $g(0) = 0$ and $\lim_{u \rightarrow \infty} g(u) = \lambda$. Furthermore
\begin{equation}
g'(u) = \frac{1}{\mu} \left[ \frac{u}{\lambda} + \frac{1}{\mu}\right]^{-2} > 0. 
\end{equation}
Hence $g$ is an increasing function. Note that since $u > -\frac{\lambda}{\mu}$ it follows that $g$ is well defined on its domain of $u \geq 0$.

Since an approximation is required for $g(u)$ it is worth examining the construction of a Taylor series for it. By successive differentiation of $g$ it can be shown that its $n$th derivative is 
\begin{equation}
g^{(n)}(u) = \frac{ (-1)^{n+1} n!}{\mu \lambda^{n-1}} \left[  \frac{u}{\lambda}  + \frac{1}{\mu}\right]^{-n-1}. \label{gfun3}
\end{equation}

One can now proceed to produce a Taylor series expansion for $g(u)$; expanding around $u=0$ tends to not be as advantageous as the choice of $u=1$.
Therefore, the relevant Taylor series is given by
\begin{equation}
g(u) = g(1) + g(1) \frac{\lambda}{\mu} \sum_{k=1}^\infty (-1)^{k+1} \left( \frac{ g(1)}{\lambda}\right)^k (u-1)^k, \label{gfunc4}
\end{equation}
where 
\begin{equation}
g(1) = \left[ \frac{1}{\lambda} + \frac{1}{\mu}\right]^{-1}. \label{gfunc5}
\end{equation}

Applying a polynomial approximation of order 2 or higher, based upon \eqref{gfunc4}, to \eqref{gfun2}, will result in the integral in \eqref{cdf1} not simplifying to a Pareto type distribution function, due to the complexity of the resulting integral. In order to achieve a Pareto approximation it is necessary to apply a linearlisation to $g(u)$. Taking a linear approximation, based upon \eqref{gfunc4}, and using the definition \eqref{gfunc5}, it can be shown that
\begin{equation}
g(u) \approx g(1)^2 \left[ \frac{u}{\mu} + \frac{1}{\lambda}\right]. \label{gfunc6}
\end{equation}

Inspection of \eqref{gfunc4} shows that the linear approximation will be valid provided $\frac{g(1)}{\lambda} << 1$, or equivalently, $\mu << \lambda + \mu$, which is requiring $\lambda$ to be large. When this condition is met, the term $g(1)\frac{\lambda}{\mu} \approx 1$.

When \eqref{gfunc6} is applied to \eqref{cdf1}, it becomes possible to evaluate the integral and show that the distribution is approximately
\begin{equation}
F_{\Z(t)} \approx 1 - \left[ \frac{\beta}{\beta + \frac{g^2(1) t}{\mu}}\right]^\alpha e^{-t\frac{g^2(1)}{\lambda}}.  \label{gfunc7}
\end{equation}
In view of \eqref{gfunc7} observe that $\frac{g^2(1)}{\lambda} = \frac{\lambda \mu^2}{[\lambda + \mu]^2} \approx O\left(\frac{1}{\lambda}\right)$.
Hence for large $\lambda$ and for fixed $t>0$, $e^{-t \frac{ g^2(1)}{\lambda}} \approx 1$. Consequently, for large $\lambda$ it follows that the distribution function \eqref{cdf1} is approximately
\begin{equation}
F_{\Z(t)} \approx 1 - \left[ \frac{\beta}{\beta + \frac{g^2(1) t}{\mu}}\right]^\alpha 
= 1 - \left[ \frac{\beta}{\beta + \frac{\lambda^2 \mu t}{\left[\lambda + \mu\right]^2}}\right]^\alpha. \label{finalcdf}
\end{equation}

Comparing the distribution function \eqref{finalcdf} with \eqref{parcdf} one concludes that in the case where $\lambda$ is larger significantly 
than $\mu$, the linearlisation is valid and the distribution function \eqref{cdf1} is approximately Pareto distributed, with shape parameter $\alpha$ and 
scale parameter $\frac{ [\lambda + \mu]^2}{\lambda^2 \mu}\beta$. Note that  as $\lambda \rightarrow \infty$,
\begin{equation}
F_{\Z(t)} \longrightarrow 1 - \left[ \frac{\beta}{\beta + \mu t}\right]^\alpha. \label{finalcdf2}
\end{equation}

This result appears to suggests that there are cases where the signal plus clutter distribution can be approximated by a Pareto model.
However, note that $\lambda$ is the reciprocal of the target power. Hence, in cases where $\lambda$ is large, the target power is very small.
The very nature of the clutter being modelled by a Pareto distribution is that it is extremely difficult to detect the presence of a low powered target 
in spiky X-band clutter \cite{weinbergbook}.  Hence, from a practical perspective, the conditions under which the approximate Pareto distribution is produced for \eqref{cdf1} are not really achievable or useful.

\section{Conclusion}
It was shown that the distribution of signal plus clutter could be approximated by a Pareto distribution, under the condition that the target model has very small power relative to the clutter level. Hence, although the approximation is valid, the conditions under which it is achieved are not feasible, since a low power signal in spiky clutter will be saturated by the clutter.


\begin{thebibliography}{}


\bibitem{finn}
Finn, H. M., Johnson, R. S. (1968).
Adaptive Detection Model with Threshold Control as a Function of Spatially Sampled Clutter-Level Estimates.
RCA Review, 29, 414-464.

\bibitem{weinbergbook}
Weinberg, G. V. (2017).
Radar Detection Theory of Sliding Window Processes. CRC Press,  New York.


    

\bibitem{siddiq}
Siddiq, K., Irshad, M. (2009).
Analysis of the Cell Averaging CFAR in Weibull Background using a Distributional Approximation.
2nd International Conference on Computer, Control and Communication.

\bibitem{dongy}
Dong, Y.  (2006).
Distribution of X-Band High Resolution and High Grazing Angle Sea Clutter.
Defence Science and Technology Organisation Research Report.

\bibitem{persson}
Persson, B. (2017).
Radar Target Modeling using In-Flight Radar Cross Section Measurements.
Journal of Aircraft, 54, 284-291.

\bibitem{weinberg13}
Weinberg, G. V. (2013).
Constant False Alarm Rate Detectors for Pareto Clutter Models.
IET Radar, Sonar and Navigation, 7, 153-163.




\end{thebibliography}
\end{document}